\documentclass{article}

\usepackage[english]{babel}

\usepackage[letterpaper,top=2cm,bottom=2cm,left=3cm,right=3cm,marginparwidth=1.75cm]{geometry}

\usepackage{amsmath}
\usepackage{graphicx}
\usepackage[colorlinks=true, allcolors=blue]{hyperref}

\usepackage[utf8]{inputenc} % allow utf-8 input
\usepackage[T1]{fontenc}    % use 8-bit T1 fonts
\usepackage{url}            % simple URL typesetting
\usepackage{amsfonts}       % blackboard math symbols
\usepackage{nicefrac}       % compact symbols for 1/2, etc.
\usepackage{microtype}      % microtypography
\usepackage{graphicx}
\usepackage{subcaption}
\usepackage{booktabs} % for professional tables
\usepackage{hyperref}
\usepackage{wrapfig}
\usepackage{multirow}
\usepackage{xcolor}
\usepackage{amsmath}
\usepackage{cite}
\usepackage[linesnumbered,ruled,vlined]{algorithm2e}
\usepackage{paralist}
\usepackage{turnstile}

\usepackage{amsthm}
\theoremstyle{definition}
\newtheorem{definition}{Definition}[section]

\newcommand{\scheme}{NetSyn}
\newcommand{\schemegen}{Genesys}
\newcommand{\vocab}{\ensuremath{\Sigma_{\mathit{DSL}}}}

\title{Learning-Based Automatic Synthesis of Software Code and Configuration}
\author{Shantanu Mandal \\ 
Texas A\&M University \\
shanto@tamu.edu}
\date{}

\begin{document}
\maketitle

\begin{abstract}
Increasing demands in software industry and scarcity of software engineers motivates researchers and practitioners to automate the process of software generation and configuration. Large scale automatic software generation and configuration is a very complex and challenging task. In this proposal, we set out to investigate this problem by breaking down automatic software generation and configuration into two different tasks.
%However, this large problem can be broken down in several different smaller tasks. We propose to explore two of the main challenging tasks of automatic program synthesis. 
In first task, %as software generation, 
we propose to synthesize software 
%also known as programs 
automatically with input output specifications. This task is further broken down into two sub-tasks. The first sub-task is about synthesizing programs with a genetic algorithm which is driven by a neural network based fitness function trained with program traces and specifications. For the second sub-task, we formulate program synthesis as a continuous optimization problem and synthesize programs with covariance matrix adaption evolutionary strategy (a state-of-the-art continuous optimization method). Finally, for the second task, 
%as software configuration, 
we propose to synthesize configurations of large scale software from different input files (e.g. software manuals, configurations files, online blogs, etc.) 
%from large scale software base 
using a sequence-to-sequence deep learning mechanism. \newline
\end{abstract}

\section{Introduction}
Automation of software generation and configuration is a well studied research problem. However, due to the complexity of the problem, there is a wide gap to fill to improve the process of automatic generation of software code and configuration. Recently, there has been notable progress in machine learning for different complex tasks. In search related problems, machine learning is heavily used due to it's incredible pattern recognition characteristics from past trained data. The trained model then helps to prune the search space significantly. Since automatic software generation is a search problem, we integrate machine learning-based approach to tackle the issue as our first task. For the second task, we formulate software configuration synthesis as a sequence-to-sequence natural language-based synthesis problem to extract configuration specifications.%\newline

%In this paragraph 
We first formulate the problem of our first task.
%, known as {\em Program Synthesis}. 
Let $S^{t} = \{(I_j, O^t_j)\}^m_{j=1}$ be a set of $m$ input-output pairs, such that the output $O_j^t$ is obtained by executing the program $P^t$ on the input $I_j$. 
Inherently, the set $S^{t}$ of input-output examples describes the behavior of the program $P^t$.
We would like to synthesize a program $P^{t'}$ that recovers the same functionality of $P^t$. However, $P^t$ is usually unknown, and we are left with the set $S^{t}$, which was obtained by running $P^t$.
Based on this assumption, we define equivalency between two programs as follows:
\begin{definition}[Program Equivalency]
\label{def:equiv}
Programs $P^a$ and $P^b$ are equivalent under the set $S = \{(I_j, O_j)\}^m_{j=1}$ of input-output examples if and only if 
$P^a(I_j) = P^b(I_j) = O_j$, for $1 \le j \le m$. We denote the equivalency by $P^a \equiv_{S} P^b$.
\end{definition}
Definition \ref{def:equiv} suggests that to obtain a program equivalent to $P^t$, we need to synthesize a program that is consistent with the set $S^{t}$.
%Then, our goal becomes to synthesize a program $P^{t'}$ that is consistent with the given set $S^{t}$ of input-output examples.
Therefore, our goal is to find a program $P^{t'}$ that is equivalent to the target program $P^t$ (which was used to generate $S^t$), i.e., $P^{t'} \equiv_{S^{t}} P^t$.
This task is known as {\em Inductive Program Synthesis} (IPS).
As suggested by \cite{deepcoder}, a machine learning-based solution to the IPS problem requires definition of two components. First, we need a programming language that defines the domain of valid programs. 
%Consequently, this language enforces a definition of the domain for the input and output samples. 
Second, we need a method to search over the program domain. The search method sweeps over the program domain to find $P^{t'}$ that satisfies the equivalency property. %\newline

We divide the first research task into two sub-tasks. 
In the first sub-task, we set out to use an off-the-self genetic algorithm for program synthesis. A genetic algorithm requires a ranking function to rank all the  solutions so that the best ones can be chosen to analyze further. 
%We use genetic algorithm for the search process and 
We propose to use a trained neural network as a ranking function. Note that a \emph{genetic algorithm} (GA) is a machine learning technique that attempts to solve a problem from a pool of candidate solutions. These generated candidates are iteratively evolved and mutated and selected for survival based on ranking function, often referred to as the \emph{fitness function}. Fitness functions are usually hand-crafted heuristics that grade the approximate correctness of candidate solutions such that those that are closer to being correct are more likely to appear in subsequent generations.
%%% can be a paragraph
% In the context of program synthesis, candidate solutions are programs, initially random but evolving over time to get closer to a program satisfying the input specification.  Yet, to guide that evolution, it is particularly difficult to design an effective fitness function for a GA-based program synthesis system.  The fitness function is given a candidate program and the input specification (e.g., input-output examples) and from those, must estimate how close that candidate program is to satisfying the specification.  However, we know that a program having only a single mistake may produce output that in no \emph{obvious} way resembles the correct output.  
% That is why, one of the most frequently used fitness functions (i.e., edit-distance between outputs) in this domain~\cite{aip, stackgp, lgp, allgp} will in many cases give wildly wrong estimates of candidate program correctness.
% Thus, it is clear that designing effective fitness functions for program synthesis is difficult.
Designing simple and effective fitness functions is a unique challenge for GA.
Despite many successful applications of GA, it still remains an open challenge to automate
the generation of such fitness functions.
An impediment to this goal is that fitness function complexity tends to increase proportionally with the problem being solved. With program synthesis being a particularly complex problem, generating a {\em good} fitness function is a daunting challenge. To solve this, we propose to explore an approach to automatically generate a fitness function by representing their structure with a neural network. %\newline 

For the second sub-task, we formulate the searching process of program synthesis from discrete domain to continuous domain and propose to solve it as a continuous optimization problem. Suppose a program $P$, consisting of $l$ tokens (instructions or functions), is denoted by $P=\left<P_1, ..., P_l\right>$, where $P_i$ represents the $i$-th token for $1\le i\le l$.
$P_i$ can be any token from the set of all possible tokens, $\vocab$. Instead of a discrete search process, we set out to investigate if an approach can be developed to map $P$ into continuous parameters that does not require any learned encoding
%any learning model 
for the formulation.
%$P$ can be mapped into continuous parameters independent of any learning model. 
%We are motivated, in part, by the common consensus that continuous parameters are easier to optimize\jt{citation needed}.  
Towards that end, we propose a {\em novel} formulation, where $P$ can be expressed as
%e aim at formulating the program as 
$P=\left<f_1(\cdot), f_2(\cdot), \dots, f_M(\cdot)\right>$. Here, each $f_i(\cdot)$, for $1\le i\le M$, is a function that takes a number of continuous parameters as inputs and maps them into some tokens in $P$. With this formulation, an error function that compares the output produced by $P$ (for a specified input) with the specified output, essentially becomes a function of continuous parameters. Therefore, the problem of program synthesis amounts to minimizing the error function. In other words, program synthesis becomes a continuous optimization problem where the goal is to minimize the error function. We propose to solve this continuous optimization problem using Covariance Matrix Adaptation Evolution Strategy (CMA-ES). %We choose CMA-ES because 
Commonly used error functions for comparing program outputs, such as edit or Manhattan distance~\cite{netsyn, aip, stackgp}, are non-smooth and ill-conditioned, i.e., a small change in the input can produce a large error. Therefore, CMA-ES is perfectly suited to solve such cases.
%For such cases, CMA-ES is considered a powerful black-box optimization technique. 
It is a 
%A common error function used for comparing program outputs is edit or Manhattan distance~\cite{netsyn, aip, pushgp}. Since this error function is %non-differentiable and 
%non-smooth and ill-conditioned (i.e., a small change in input can produce a large error),
%we propose to use a Covariance Matrix Adaptation Evolution Strategy (CMA-ES) to solve the continuous optimization problem. CMA-ES is a 
stochastic derivative-free algorithm for difficult (e.g., non-convex, ill-conditioned, multi-modal, rugged, noisy, etc.) optimization problems and considered as one of the most advanced %evolutionary 
optimization algorithms. %\newline

%In this paragraph 
Finally, we formulate the problem of our second task - configuration synthesis. Every large scale software takes some inputs to configure them. These inputs are crucial and can vary across different software. Configuring software with invalid inputs can lead to software failures. Sometimes these inputs are not checked inside the software rather are specified in some software manuals or other places (such as readme files, online blogs, etc.). 
Based on the use cases, admins configure the software. However, invalid configurations can lead to software failures. Thus, it is important to automatically generate specification sets from configurations to prevent software misconfiguration failure. Due to the unstructured format of the input files, it is hard to extract configuration constraints automatically. To tackle this problem, we propose a system that uses natural language-based machine learning model to automatically extract configuration constraints from different input files.

\section{Task 1: Automatic Software Synthesis}

In Task 1, we propose to synthesize programs automatically that satisfy some specifications. We approach this problem in two different ways. We use discrete search process and a continuous search process. They are described in the following sections as sub-task 1 and 2. These sub-tasks are named as NetSyn\cite{netsyn} and Genesys\cite{genesys}.

\subsection{Sub-Task 1: NetSyn}
% -------------------  Design -------------------------

% Here, we describe our solution to IPS  in more detail, including the choices and novelties for each of the proposed components. We name our solution \scheme{} as it is based on neural networks for program synthesis. 
%In this section, we describe \scheme's design as illustrated in Figure~\ref{fig-overview}. We begin with a brief overview of the three main phases of \scheme\ and subsequently discuss some of the novel elements of these phases thereafter.

Here, we describe our proposed solution to synthesize programs with a genetic algorithm (GA) and a trained neural network used as a fitness function in GA. We name our solution as \scheme. In the following sections, we will first describe the domain specific language that we use, the GA search process, how we learn the fitness function, and finally, the auxiliary local neighborhood search.  

\subsubsection{Domain Specific Language}
\label{sec-domain}
As \scheme{}'s programming language, we choose a domain specific language (DSL) constructed specifically for it. This choice allows us to constrain the program space by restricting the operations used by our solution.
% based on previous work
\scheme's DSL follows the DeepCoder's DSL \cite{deepcoder}, which was inspired by SQL and LINQ~\cite{Kulkarni:2007}. 
% Data types
The only data types in the language are \emph{(i)} integers and  \emph{(ii)} lists of integers.
% What functions
The DSL contains 41 functions, each taking one or two arguments and returning one output. Many of these functions include operations for list manipulation. Likewise, some operations also require lambda functions. There is no explicit control flow (conditionals or looping) in the DSL. However, several of the operations are high-level functions and are implemented using such control flow structures. A full description of the DSL can be found in the supplementary material. 
% what is a program
With these data types and operations, we define a program $P$ as a sequence of functions. Table \ref{table:example} presents an example of a program of 4 instructions with an input and respective output.

%\scheme's DSL is similar to that of DeepCoder~\cite{deepcoder} and is inspired by SQL and LINQ~\cite{Kulkarni:2007}. The only data types in \scheme's DSL are \emph{(i)} integers and  \emph{(ii)} lists of integers. A program in this DSL is a sequence of high level functions, each taking one or two arguments and returning one output. Similar to, but not precisely like, Halide's language design, there is no explicit control flow (conditionals or looping) in our DSL~\cite{Mullapudi:2016:siggraph, Adams:2019:siggraph}.  

Arguments to functions are not specified via named variables. Instead, each function uses the output of the previously executed function that  produces the type of output that is used as the input to the next function. The first function of each program uses the provided input $I$. If $I$ has a type mismatch, default values are used (i.e., 0 for integers and an empty list for a list of integers). The final output of a programs is the output of its last function. 
%\jt{Should we add an additional example of input that has a mismatch within the example?}

\begin{table}[htpb]
  %\scriptsize
  %\parbox{0.3\linewidth}{
%   \begin{minipage}{\columnwidth}
%\vspace{-0.2cm}
  \caption{An example program of length 4 with an input and corresponding output.}
  \label{table:example}
   \centering
    \scalebox{1}{
  %\vskip 0.15in
%  \begin{center}
%  \begin{footnotesize}
  %\begin{sc}
  \begin{tabular}{ll}
  \hline
%   a $\leftarrow$ [int] & Input: \\
%   b $\leftarrow$ \textsc{Filter ($>$0)} a & [-2, 10, 3, -4, 5, 2] \\
%   c $\leftarrow$ \textsc{Map (*2)} b &  \\
%   d $\leftarrow$ \textsc{Sort} c & Output:\\
%   e $\leftarrow$ \textsc{Reverse} d & [20, 10, 6, 4] \\\hline
   [int] & Input: \\
   \textsc{Filter ($>$0)}  & $[-2, 10, 3, -4, 5, 2]$ \\
   \textsc{Map (*2)}  &  \\
   \textsc{Sort}  & Output:\\
   \textsc{Reverse}  & $[20, 10, 6, 4]$ \\\hline

  \end{tabular} }
  %\vspace{1cm}
  %}
%  \end{minipage}
%\end{sc}
%%\end{footnotesize}
%\end{center}
%\vspace{0.1cm}
%\vspace{-0.2cm}
\end{table}

%When taken as a whole, \scheme's DSL has a novelty and amenability to genetic algorithms. 
As a whole, \scheme's DSL is novel and amenable to genetic algorithms. The language is defined such that all possible programs are \emph{valid by construction}. This makes the whole program space valid and is important to facilitate the search of programs by any learning method. 
In particular, this is very useful in evolutionary process in genetic algorithms. When genetic crossover occurs between two programs or mutation occurs within a single program, the resulting program will \emph{always} be valid. This eliminates the need for pruning to identify valid programs.

\begin{figure*}[htpb]
 %   \vskip 0.2in
    \begin{center}
        %\centerline{\includegraphics[trim=0 10 0 230, clip, width=0.8\textwidth]{./FIGS/Overview_new}}
        \includegraphics[width=0.8\textwidth]{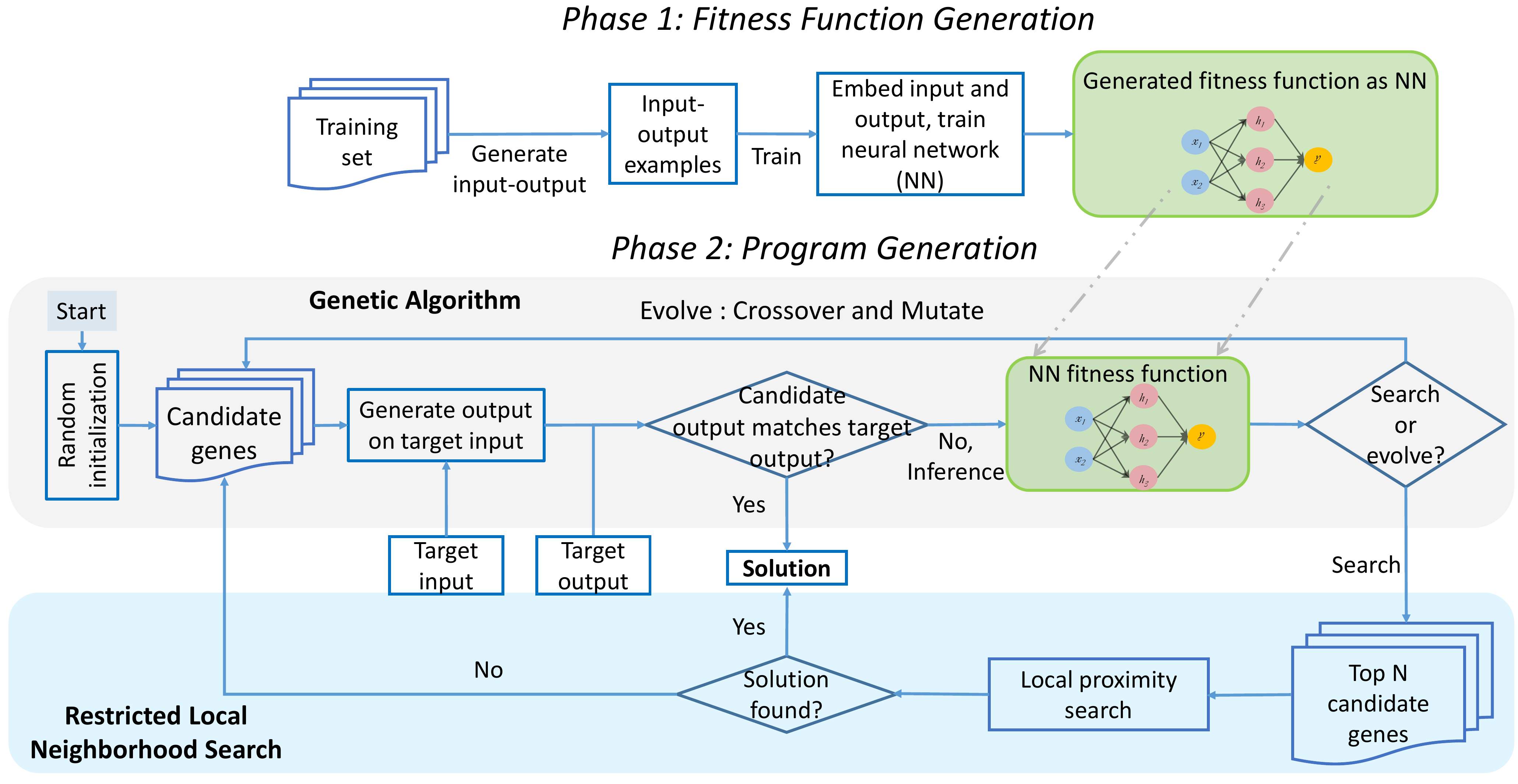}
        \caption{Overview of \scheme. Phase 1 automates the fitness function generation by training a neural network on a corpus of example programs and their inputs and outputs. Phase 2 finds the target program for a given input-output example using the trained neural network as a fitness function in a genetic algorithm.}
        \label{fig-overview}
    \end{center}
 %   \vspace{-0.4cm}
\end{figure*}

\subsubsection{Search Process}
%\scheme{}'s search method is based on a genetic algorithm \jt{\cite{AA}}. 
%Given a population of genes $\Phi^j$, the algorithm evolves them into a new generation $\Phi^{j+1}$. The search repeats iteratively until a solution is found. In our case, a solution is a gene that describes a program equivalent to the target program $P^t$ under the set $S^t$ of input-output samples.
%We define a gene to be a program in the DSL.
%General (high-level) description
\scheme\ synthesizes a program by searching the program space with a genetic algorithm-based method~\cite{globalopt}. It does this by creating a population of random genes (i.e., candidate programs) of a given length $L$ and uses a learned neural network-based fitness function (NN-FF) to estimate the fitness of each gene. Higher graded genes are preferentially selected for crossover and mutation to produce the next generation of genes. In general, \scheme\ uses this process to evolve the genes from one generation to the next until it discovers a correct candidate program as verified by the input-output examples. From time to time, \scheme\ takes the top $N$ scoring genes from the population, determines their neighborhoods, and looks for the target program using a local proximity search. If a correctly generated program is not found within the neighborhoods, the evolutionary process resumes. Figure~\ref{fig-overview} summarizes the \scheme{}'s search process. 

% Gene encoding for each program
%\subsubsection{Gene Encoding}
%\label{sec-encoding
We use a value encoding approach for each gene. A gene $\zeta$ is represented as a sequence of values from \vocab{}, the set of functions. Formally, a gene $\zeta = \left(f_1,\dots,f_i,\dots,f_L \right)$, where $f_i \in \vocab{}$. Practically, each $f_i$ contains an identifier (or index) corresponding to one of the DSL functions. The encoding scheme satisfies a one-to-one match between programs and genes. %Additionally, this encoding simplifies the crossover and mutation implementation by eliminating the potential of invalid gene construction. \jt{The last sentence seems out of place as we claimed that all programs are valid from construction. Please confirm that I'm right, and remove the sentence.}

%Initialization + evolution process
The search process begins with a set $\Phi^0$ of $|\Phi^0|=T$ randomly generated programs.
If a program equivalent to the target program $P^t$ is found, the search process stops.
Otherwise, the genes are ranked using a learned NN-FF. A small percentage (e.g., 20\%) of the top graded genes in $\Phi^j$ are passed in an unmodified fashion to the next generation $\Phi^{j+1}$ for the next evolutionary phase. This guarantees that some of the top graded genes are identically preserved, aiding in forward progress guarantees. The remaining genes of the new generation $\Phi^{j+1}$ are created through crossover or mutation with some probability. For crossover, two genes from $\Phi^j$ are selected using the Roulette Wheel algorithm with the crossover point selected randomly~\cite{Goldberg:1989:awlp}. 
%For mutation, one gene is mutated to some other 
For mutation, one gene is Roulette Wheel selected and the mutation point $k$ in that gene is selected based on the same learned NN-FF.
%randomly. 
The selected value $z_k$ is mutated to some other random value $z'$ such that $z' \in \vocab{}$ and $z'\ne z_k$. 

Crossovers and mutations can occasionally lead to a new gene with dead code. To address this issue, we eliminate dead code. Dead code elimination (DCE)
is a classic compiler technique to remove code from a program that has no effect on the program's output~\cite{Debray:2000:acm}. Dead code is possible in our list DSL if the output of a statement is never used. We implemented DCE in \scheme\ by tracking the input/output dependencies between statements and eliminating those statements whose outputs are never used.  \scheme\ uses DCE during candidate program generation and during crossover/mutation to ensure that the effective length of the program is not less than the target program length due to the presence of dead code.
If dead code is present, we repeat crossover and mutation until a gene without dead code is produced. 

%To begin the GA process, \scheme\ initializes the population with randomly generated genes. Unless a correct program is found (i.e., an exact match of candidate program output compared to target output), the remaining genes are graded using the FF neural network model described in Phase 1. A small percentage (e.g., 20\%) of top graded genes are moved into the next evolutionary phase in an unmodified fashion. This guarantees some of the top graded genes are identically preserved, aiding in forward progress guarantees. 

%The remaining genes are created through crossover or mutation with some probability. For crossover, two genes are selected using the Roulette Wheel algorithm with the crossover point selected pseudo-randomly~\cite{Goldberg:1989:awlp}. For mutation, one gene is Roulette Wheel selected and the mutation point in that gene is selected pseudo-randomly. The selected value $C$ is mutated to some other pseudo-random value $C'$ such that $C'\in \{x\}^{|\Sigma_{\mathit{DSL}}|}_{x=1}$ and $C'\ne C$. Crossover and mutation can sometimes lead to a new gene with dead code. \scheme\ checks for dead code (Section ~\ref{dead_code_elimination}). If dead code is present, \scheme\ repeats crossover and mutation until a gene without dead code is produced.  

\begin{figure*}[htpb]
    \centering
    
    \begin{subfigure}{\textwidth}
        \includegraphics[width=\textwidth]{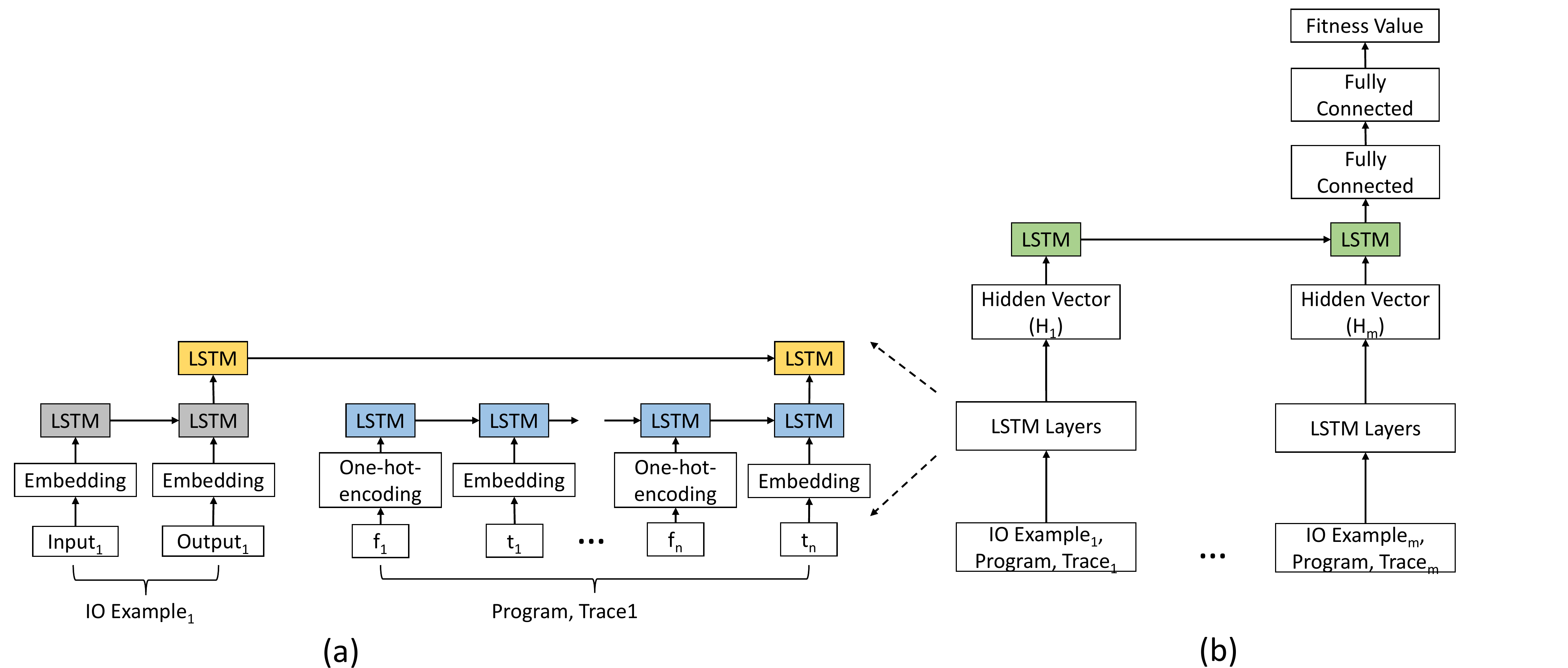}
    \end{subfigure}
%    \quad
%    \begin{subfigure}{0.25\textwidth}
%        \includegraphics[width=0.95\textwidth]{./FIGS/NS.pdf}
%    \end{subfigure}

    \caption{Neural network fitness function for (a) single and (b) multiple IO examples. In each figure, layers of LSTM encoders are used to combine multiple inputs into hidden vectors for the next layer. Final fitness score is produced by the fully connected layer.}
    %Examples of neighborhood using (c) BFS- and (d) DFS-based approach.}
    \label{fig:model}
    %\vspace{-0.3cm}
\end{figure*}

\subsubsection{Learning the Fitness Function}
\label{sec-fitness}

% What a FF is, how it is used, and the handcrafting challenge
Evolving the population of genes in a genetic algorithm requires a fitness function to rank the fitness (quality) of genes based on the problem being solved. Ideally, a fitness function should measure how close a gene is to the solution. Namely, it should measure how close a candidate program is to an equivalent of $P^t$ under $S^t$. Finding a good fitness function is of great importance to reduce the
number of steps in reaching the solution and directing the algorithm in the right direction so that 
%size of the search domain and to make the 
genetic algorithm are more likely to find $P^t$. 
%We don't reduce the search domain size. This space is the same. We want to reach the target in less steps.
%in less time.
%Of course, this could be accomplished if the solution would be available to us. However, this is never the case.

% Circle back on the prediction proxy novelty 
{\bf Intuition:} A fitness function, often, is handcrafted to approximate some ideal function that is impossible (due to incomplete knowledge about the solution) or too computationally intensive to implement in practice. For example, if we knew $P^t$ beforehand, we could have designed an ideal fitness function that compares a candidate program with $P^t$ and calculates some metric of closeness (e.g., edit distance, the number of common functions etc.) as the fitness score. Since we do not know $P^t$, we cannot implement the ideal fitness function.
%This function ends up serving as an approximation proxy to the ideal one. 
Instead, in this work, we propose to approximate the ideal fitness function by learning it from training data (generated from a number of known programs).
%Instead, in this work, we propose to learn the fitness function itself such that we can predict the fitness score of genes. 
For this purpose, we use a neural network model. We train it with the goal of predicting the values of an ideal fitness function.  We call such an ideal fitness function (that would always give the correct answer with respect to the actual solution) the \emph{oracle} fitness function as it is impossible to achieve in practice merely by examining input-output examples.
%\taa{Proof for the preceding statement?}
In this case, our models will not be able to approach the 100\% accuracy of the \emph{oracle} but rather will still have sufficiently high enough accuracy to allow the genetic algorithm to make forward progress.
%Therefore, this model acts as a \emph{prediction} proxy. 
Also, we note that the trained model needs to generalize to predict for any unavailable solution and not a single specific target case.

% We follow ideas from works that have explored the automation of fitness functions using neural networks for approximating a known mathematical model. For example, Matos Dias et al.~\cite{Dias:2014:cejor} automated them for IMRT beam angle optimization, while Khuntia et al.~\cite{Bonomali:2005:motl} used them for rectangular microstrip antenna design automation. In contrast, our work is fundamentally different in that we use a large corpus of program metadata to train our models to predict 
% %the likelihood that a given, incorrect solution will eventually converge to become a program generating the correct output. 
% how close a given, incorrect solution could be from an {\em unknown} correct solution (that will generate the correct output).
% In other words, we propose to automate the generation of fitness functions using big data learning.
% To the best of our knowledge, \scheme\ is the {\em first} proposal for automation of fitness functions in genetic algorithms. In this paper, we demonstrate this idea using MP as the use case.
%of its kind automated \emph{prediction} fitness function for big data learning. 

% the training process 
% inference with NN-FF
Given the input-output samples $S^t=\left\{\left(I_j, O_j^t\right)\right\}_j$ of the target program $P^t$ and an ideal fitness function $fit(\cdot)$, we would like a model that predicts the fitness value $fit(\zeta, P^t)$ for a gene $\zeta$. In practice, our model predicts the values of $fit(\cdot)$ from input-output samples in $S^t$ and from execution traces of the program $P^\zeta$ (corresponding to $\zeta$) by running with those inputs. Intuitively, execution traces provide insights of whether the program $P^\zeta$ is on the right track.
%samples with the output of $\zeta$:  $S^\zeta=\left\{\left(I_j, O_j^\zeta\right)\right\}_j$, with $O_j^\zeta=\zeta\left(I_j\right)$.

% the model of NN-FF, data needs
In \scheme{}, we use a neural network to model the fitness function, referred to as NN-FF. This task requires us to  generate a training dataset of programs with respective input-output samples. To train the NN-FF, we randomly generate a set of example programs, $E=\left\{P^{e_j}\right\}$, along with a set of random inputs $I^j=\{I_i^{e_j}\}$ per program $P^{e_j}$. We then execute each program $P^{e_j}$ in $E$ with its corresponding input set $I^j$ to calculate the output set $O^j$. Additionally, for each $P^{e_j}$ in $E$, we randomly generate another program $P^{r_j}=(f^{r_j}_{1}, f^{r_j}_{2}, ..., f^{r_j}_{n})$, where $f^{r_j}_{k}$ is a function from the DSL i.e., $f^{r_j}_{k} \in~\vocab{}$. 
%\jt{We used $z_i$ in the previous section. Also it is easy to confuse $f$ and $f^{r_j}$.}
We apply the previously generated input $I_i^{e_j}$ to $P^r_j$ to get an execution trace, $T^{rj}_i=(t^{rj}_{i1}, t^{rj}_{i2}, ..., t^{rj}_{in})$, where $t^{rj}_{ik}=f^{rj}_{k}(t^{rj}_{i(k-1)})$ with $t^{rj}_{i1}=f^{rj}_{1}(I_i^{e_j})$ and $t^{rj}_{in}=f^{rj}_{n}(t^{rj}_{i(n-1)})=P^{r_j}(I_i^{e_j})$. Thus, the input set $I^j=\{I^{e_j}_i\}$ of the program $P^{e_j}$ produces a set of traces $T^j=\{T^{r_j}_i\}$ from the program $P^{r_j}$.
%We apply the previously generated input $I_i^{e_j}$ to $P^r$ to produce the output $O_i^r=P^r(I_i^{e_j})$. 
We then compare the programs $P^{r_j}$ and $P^{e_j}$ 
%and their outputs \taa{is this part about outputs correct?  don't think so...only the programs are compared right?} 
to calculate the fitness value and use it as an example to train the neural network. 

In \scheme, the inputs of NN-FF consist of input-output examples, generated programs, and their execution traces. Let us consider 
%the case of 
a single input-output example, $(I^{e_j}_i, O^{e_j}_i)$. Let us assume that $P^{e_j}$ is the target program that \scheme\ attempts to generate and in the process, it generates $P^{r_j}$ as a potential equivalent. NN-FF uses $(I^{e_j}_i$, $O^{e_j}_i)$, and $\{(f^{r_j}_{k}, t^{r_j}_{ik})\}$ as inputs for this example. Each of $(I^{e_j}_i$, $O^{e_j}_i)$, and $t^{r_j}_{ik}$ are passed through an embedding layer followed by an LSTM encoder. $f^{r_j}_{k}$ is passed as a one-hot-encoding vector. 
%\sh{Following text needs to be updated to accommodate the new model without single fitness value column} 
Figure~\ref{fig:model}(a) shows the details of how 
%NN-FF architecture for 
a single input-output example is processed. Two layers of LSTM encoders combines the vectors to produce a single vector, $H^j_i$.
%, which is then processed through fully connected layers to predict the fitness value. 
In order to handle a set of input-output examples, $\{(I^{e_j}_i, O^{e_j}_
i)\}$, a set of execution traces, $T^j=\{T^{r_j}_i\}$, is collected from a single generated program, $P^{r_j}$. Each input-output example, $(I^{e_j}_i, O^{e_j}_i)$, along with the corresponding execution trace produces a single vector, $H^j_i$. An LSTM encoder combines such vectors to produce a single vector, which is then processed by fully connected layers to predict the fitness value (Figure~\ref{fig:model}(b)).

{\bf Example:} To illustrate, suppose the program in Table~\ref{table:example} is in $E$. Let us assume that $P^{r_j}$ is another program \{\textsc{[int], Filter ($>$0), Map (*2), Reverse, Drop (2)}\}. If we use the input in Table~\ref{table:example} (i.e., $[-2, 10, 3, -4, 5, 2]$) with $P^{r_j}$, the execution trace is $\left\{[10, 3, 5, 2]\right.$, $[20, 6, 10, 4]$, $[4, 10, 6, 20]$, $\left.[6, 20]\right\}$. 
So, the input of NN-FF is \{$[-2, 10, 3, -4, 5, 2], [20, 10, 6, 4]$, $Filter_{v}$, $[10, 3, 5, 2]$, $Map_{v}$, $[20, 6, 10, 4]$, $Reverse_v$, $[4, 10, 6, 20]$, $Drop_v$, $[6, 20]$\}. $f_v$ indicates the value corresponding to the function $f$.

%the output $O^r$ is [6, 20]. In case of $IO$ model, the input for NN-FF is \{[-2, 10, 3, -4, 5, 2], [6, 20]\}. For $IO^2$ and $IO^\delta$ model, the inputs are \{[-2, 10, 3, -4, 5, 2], [20, 10, 6, 4], [-2, 10, 3, -4, 5, 2], [6, 20]\} and \{[-22, 0, -3, -8, 5, 2], [-8, -10, 3, -4, 5, 2]\} respectively.

% The specific functions we use for program synthesis: CF, LCS, FP
%We design the following three fitness metrics -- common functions, longest common subsequence, and function probability -- which we discuss next, and compare their performance in synthesizing programs in Section~\ref{sec-results}.
There are different ways to quantify how close two programs are to one another.  Each of these different methods then has an associated metric and ideal fitness value.  We investigated three such metrics -- common functions, longest common subsequence, and function probability -- which we use as the expected predicted output for the NN-FF.

{\bf Common Functions: }
\scheme\ can use the number of common functions (CF) between $P^\zeta$ and $P^t$ as a fitness value for $\zeta$. In other words, the fitness value of $\zeta$ is
%\begin{equation}
$f^{CF}_{P^t}(\zeta)=|\mathbf{elems}(P^\zeta) \cap \mathbf{elems}(P^t)|$. For the earlier example, $f^{CF}$ will be 3.
%\end{equation}
%\scheme\ constructs training data for $f^{CF}$ from $E$ as described above and feeds to a neural network for training. 
Since the output of the neural network will be an integer from 0 to $\mathbf{len}(P_t)$, the neural network can be designed as a multiclass classifier with a softmax layer as the final layer.

{\bf Longest Common Subsequence: }
As an alternative to CF, we can use longest common subsequence (LCS) between $P^\zeta$ and $P^t$. The fitness score of $\zeta$ is 
%\begin{equation}
$f^{LCS}_{P^t}(\zeta)=\mathbf{len}(\mathit{LCS}(P^\zeta, P^t))$.
%\end{equation} 
Similar to CF, training data can be constructed from $E$ which is then fed into a neural network-based multiclass classifier. For the earlier example, $f^{LCS}$ will be 2.
%Inputs of the neural network can be $IO$ or $IO^2$ or $IO^\delta$.

{\bf Function Probability: }
The work \cite{deepcoder} proposed a probability map for the functions in the DSL. %We construct an ideal fitness function using the probability map. 
Let us assume that the probability map $\mathbf{p}$ is defined as the probability of each DSL operation to be in $P^t$ given the input-output samples. Namely, $\mathbf{p} = (p_1, \dots, p_k,\dots,p_{|\vocab{}|})$ such that $p_k=Prob(\mathrm{op}_k \in \mathbf{elems}(P^t)|\{(I_j, O^t_j)\}^m_{j=1})$, where $\mathrm{op}_k$ is the $k^{th}$ operation in the DSL. 
Then, a multiclass, multilabel neural network classifier with sigmoid activation functions used in the output of the last layer can be used to predict the probability map. Training data can be constructed for the neural network using $E$. We can use the probability map to calculate the fitness score of $\zeta$ as 
%\begin{equation}
$f^{FP}_{P^t}(\zeta)=\sum_{k:\mathrm{op}_k \in \mathbf{elems}(P^\zeta)} p_k$.
%\end{equation}
%\jt{The original formula for $f^{FP}$ didn't make sense to me. Somebody, please, make sure that now makes sense or let me know what should be.}
\scheme\ also uses the probability map to guide the mutation process. For example, instead of mutating a function $z_k$ with $z'$ that is selected randomly, \scheme\ can select $z'$ using Roulette Wheel algorithm using the probability map.

%--------------------------------------------------------------------------------------------------------

\subsubsection{Local Neighborhood Search}
\label{sec-neighbor}

%\begin{figure}[tb]
%    \vskip -0.2in
%    \begin{center}
%    \centerline{
%    \includegraphics[width=0.5\columnwidth]{./FIGS/NS2}
%    }
%    \caption{Example of neighborhood for a $4$-length gene using (a) BFS- and (b) DFS-based approach.}
%    \label{fig-neighborhood}
%    \end{center}
%    \vskip -0.2in
%\end{figure}

Neighborhood search (NS) checks some candidate genes in the {\em neighborhood} of the $N$ top scoring genes from the genetic algorithm. The intuition behind NS is that if the target program $P^t$ is in that neighborhood, \scheme\ may be able to find it without relying on the genetic algorithm, which would likely result in a faster synthesis time.
%to synthesize the correct program.

% \begin{figure}[htpb]
%     \centering
    
%     \begin{subfigure}{0.25\textwidth}
%         \includegraphics[width=\textwidth]{./FIGS/NS3.pdf}
%         \caption{BFS-based}
%     \end{subfigure}
% %    \quad
%     \begin{subfigure}{0.25\textwidth}
%         \includegraphics[width=0.95\textwidth]{./FIGS/NS4.pdf}
%         \caption{DFS-based}
%     \end{subfigure}

%     \caption{Examples of neighborhood using (a) BFS- and (b) DFS-based approach. Each neighborhood constructs a set of close-by genes by systematically changing one function at a time.}
%     \label{fig:neighbor}
%     %\vspace{-0.3cm}
% \end{figure}
%
%\paragraph{Neighborhood Search Invocation}
Let us assume that \scheme{} has completed $l$ generations. Then, let $\mu_{l-w+1,l}$ denote the average fitness score of genes for the last $w$ generations (i.e., from $l-w+1$ to $l$) and $\mu_{1,l-w}$ will denote the average fitness score before the last $w$ generations (i.e., from $1$ to $l-w$). Here, $w$ is the sliding window. \scheme\ invokes NS if $\mu_{l-w+1,l} \le \mu_{1,l-w}$. The rationale is that under these conditions, the search procedure has not produced improved genes for the last $w$ generations (i.e., saturating). Therefore, it should check if the neighborhood contains any program equivalent to $P^t$.

%\vspace{-0.4cm}
%%%%%%%%%% Need to use a different package
\begin{algorithm}[htb]
%\scriptsize
\footnotesize
\KwIn{A set $G$ of top $N$ scoring genes}
\KwOut{$P^{t'}$, if found, or \emph{Not found} otherwise}
\For{Each $\zeta \in G$}{
  $NH \gets \emptyset$
  
  \For{$i \gets 1~\mathbf{to~len}(\zeta)$}{
    \For{$j \gets 1~\mathbf{to} |\Sigma_{DSL}|$}{
      $\zeta_n \gets \zeta~with~\zeta_i~replaced~with~\mathrm{op}_j$
      $~~such~that~\zeta_i \ne \mathrm{op}_j$
      
      $NH \gets NH \cup \{ \zeta_n\}$
    }
  }
  \If{there is $P^{t'}\in NH$ such that $P^{t'}\equiv_{S^t} P^t$}{
  \Return{$P^{t'}$}
  }
}
\Return{Not found}
\caption{Defines and searches neighborhood based on BFS principle}
\label{algo-neighbor}
%\vspace{-0.4cm}
\end{algorithm}
%\vspace{-0.4cm}

{\bf Neighborhood Definition: }
Algorithm~\ref{algo-neighbor} shows how to define and search a neighborhood. The algorithm is inspired by the breadth first search (BFS) method. For each top scoring gene $\zeta$, \scheme\ considers one function at a time starting from the first operation of the gene to the last one. Each selected operation is replaced with all other operations from $\vocab{}$, and inserts the resultant genes into the neighborhood set $NH$. If a program $P^{t'}$ equivalent to $P^t$ is found in $NH$, \scheme\ stops there and returns the solution. Otherwise, it continues the search and returns to the genetic algorithm. The complexity of the search is $\mathcal{O}(N\cdot \mathbf{len}(\zeta)\cdot|\vocab{}|)$, which is significantly smaller than the exponential search space used by a traditional BFS algorithm.
Similar to BFS, \scheme\ can define and search the neighborhood using an approach similar to depth first search (DFS). It is similar to Algorithm~\ref{algo-neighbor} except $i$ keeps track of depth here. After the loop in line 4 finishes, \scheme\ picks the best scoring gene from $NH$ to replace $\zeta$ before going to the next level of depth. The algorithmic complexity remains the same.

\subsection{Sub-Task 2: GeneSys}
% ----------------------- Design ----------------------
Here we propose our solution to synthesize programs with continuous optimization technique. We call this solution as \schemegen. In the following sections we will describe how we formulate the program synthesis problem as a continuous optimization problem. We use covariance matrix adaptive evolutionary technique, in short CMA-ES. In part of that we will describe different mapping schemes that we propose to incorporate the problem into CMA-ES and different restart policies. We use the same DSL described in \ref{sec-domain}.

\subsubsection{Program Synthesis as a Continuous Optimization Problem - A Novel Formulation}
\label{sec-cont-opt}
\begin{figure}[hptb]
\begin{center}
\includegraphics[width=0.45\columnwidth]{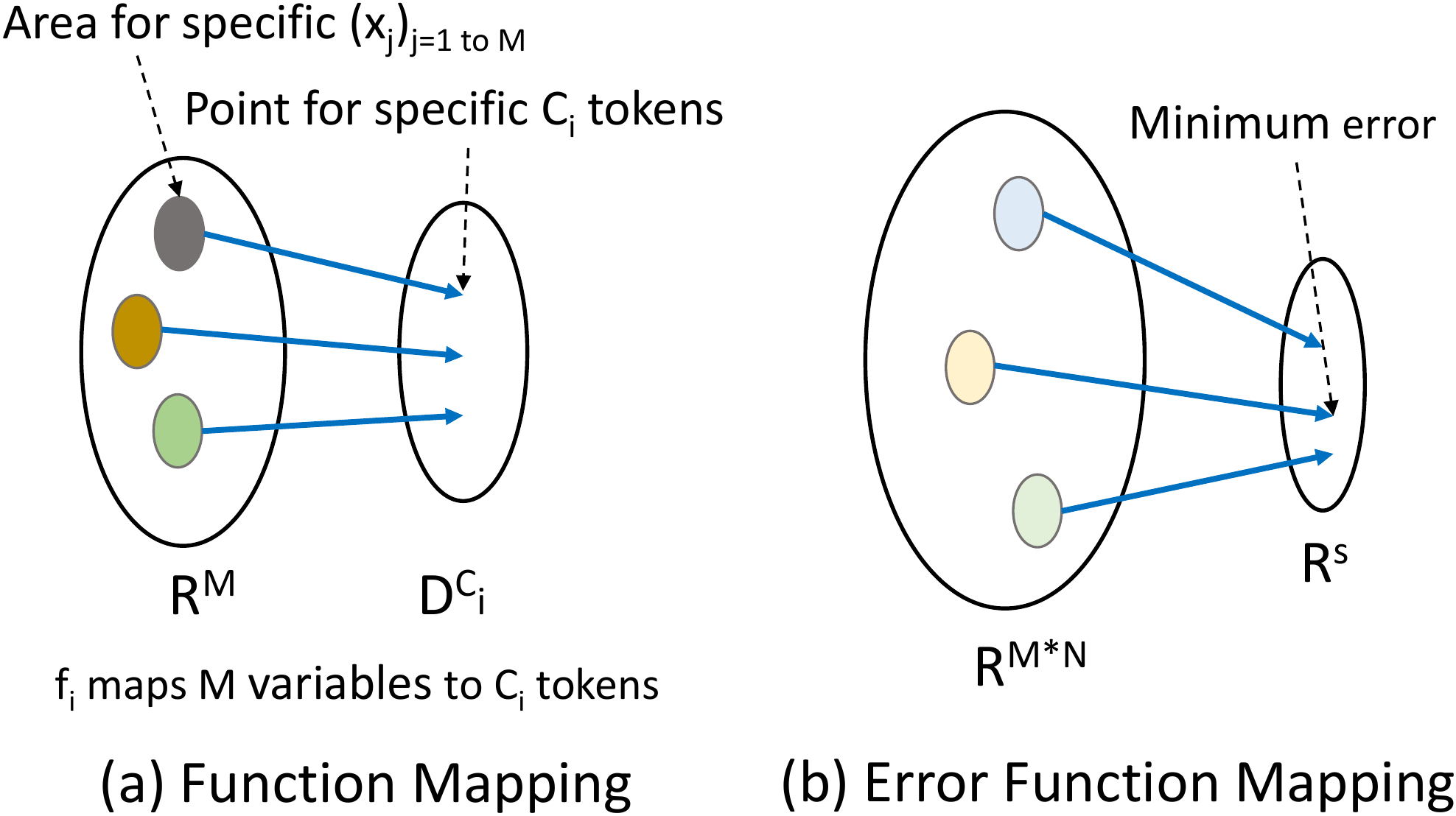}
%\vspace{-0.2cm}
\caption{Pictorial representation of how (a) each function maps the continuous parameters to DSL tokens in the program and (b) the error function maps the continuous parameters.w    }%\jt{figure is too small and hard to read and see the arrow heads.}}
\label{fig_mapping}
\end{center}
\vspace{-0.6cm}
\end{figure}

We propose to model a program as an $l$-tuple, 
%where $l$ is the length of the program and 
where 
each element of the tuple is a function of continuous parameters. 
%Being $l\ge 1$ the length of the program. 
Intuitively, each function takes one or more continuous parameters and maps them to $0$ or more DSL tokens in the program.
%With the DSL in Section~\ref{sec-domain} being the programming language, $n$ indicates the number of DSL functions in the program. 
Let us consider a program $P$. It can be expressed as $P=\left<f_1(x^{1}_1, ..., x^{1}_M), ..., f_N(x^{N}_1, ..., x^{N}_M)\right>$ where, $f_i(x^{i}_1, ..., x^{i}_M): \mathbb{R}^M\rightarrow \vocab^{C_i}$ for $1\le i\le N$, $0\le C_i\le l$, and $N,M>0$. Namely, each function $f_i$ maps $M$ continuous random variables into $C_i$ DSL tokens in the program. When $C_i=0$, it indicates the special case when $f_i$ does not map its parameters to any DSL token. In other words, $f_i$ becomes a NULL function.
%, $\vocab^{C_i}$ essentially becomes an empty set, $\emptyset$.
Figure~\ref{fig_mapping}(a) shows the mapping from the continuous variables to DSL tokens. Since the length of the program $P$ is $l$,  we can infer $\Sigma_iC_i=l$.

%\begin{figure*}[!ht]
%    \centering
%    \begin{subfigure}{0.65\columnwidth}
%        \includegraphics[width=0.8\columnwidth]{./FIGS/mapping}
%    %\vspace{-0.6cm}
%    \caption{Function Mapping \jt{Why don't we have one figure generalized to fi, Ci? Also, can we add x1,...,xM in the figure? And maybe some DSL token names in the right side?}} \label{fig_mapping_a}
%    \end{subfigure}
%    %\quad
%    \begin{subfigure}{0.33\columnwidth}
%        \includegraphics[width=0.6\columnwidth]{./FIGS/error}
%     %\vspace{-0.6cm}
%     \caption{Error Function Mapping \jt{This figure is confusing. The MxN space maps to outputs of P, and those compared to the actual outputs. Shall we add the actual outputs? Shall we add E(P(Ij),O) to the small ellipse?}}
%    \end{subfigure}
%    \caption{Pictorial representation of how (a) each function maps the continuous %parameters to DSL tokens in the program and (b) the error function maps the continuous parameters. \jt{R and D have a different math font than the text in these figures}}
%    \label{fig_mapping}
%    % \vspace{-0.4cm}
%\end{figure*}

We define an error over the given specification, $S=\{(I_j, O_j)\}$ as 
%an $s$-tuple, 
$\mathbb{E}(P, S)=\left<E(P(I_j), O_j)\right>^s_{j=1}$.
%\jt{Does the error term represents a tuple or a scalar? It doesn't seem to match the text that follows.}. 
Here, $E(P(I_j), O_j)$ could be any commonly used distance function in program synthesis such as edit distance, Manhattan distance, etc.~\cite{Jurafsky2009}. With the formulation of $P$ as an $l$-tuple, $\mathbb{E}$ becomes a function mapping $\mathbb{R}^{M\times N}$ to $\mathbb{R}^{s}$. Therefore, program synthesis becomes a continuous optimization problem where the goal is to find values of $M\times N$ continuous random variables that minimize $\mathbb{E}$ (i.e., the minimum point in the $s$-dimensional space). Figure~\ref{fig_mapping}(b) depicts this formulation. 
%\jt{How is the error defined, when the output O is a list of integers that is of different length than the expected Oj?} \jt{Also, how is the error distance E minimize when there are s error terms (one for each input-output pair)?}

%\begin{figure}[hptb]
%\begin{center}
%\includegraphics[width=0.2\columnwidth]{./FIGS/error}
%%\vspace{-0.2cm}
%\caption{Pictorial representation of how the error function maps the continuous parameters. This formulation makes the program synthesis a continuous optimization problem.}
%\label{fig_error}
%\end{center}
%\vspace{-0.4cm}
%\end{figure}

%\subsection{Program construction with \acro\ variables}
\subsubsection{Mapping Schemes}
\label{sec-sub-setup}
\begin{figure}[hptb]
\begin{center}
\includegraphics[width=1.0\columnwidth]{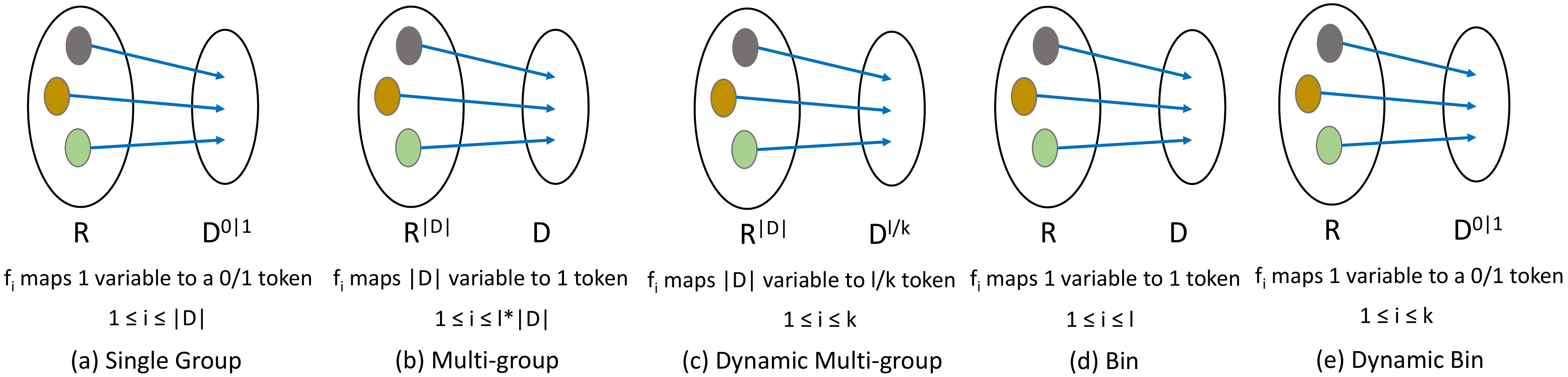}
\vspace{-0.4cm}
\caption{Representation of different mapping schemes.}
%\jt{R and D have a different math font than the text in these figures 
%$D^{0|1}$ has 2 options in the right ellipse. The second drawing appears like there were 3. Also, like in figure 1, shall we add labels for x1..xL and DSL tokens?}}
\label{fig_diff_mapping}
\end{center}
\vspace{-0.8cm}
\end{figure}

Based on the problem formulation in the prior Section, we propose a number of {\em novel mapping} schemes for \schemegen. Each scheme maps one or more continuous random variables to $0$ or more  DSL tokens of a program. \schemegen\ uses these mapping schemes to convert random variable samples to actual programs during the CMA-ES algorithm.
%Figure~\ref{fig_single_group}-\ref{fig_dyn_bin_mapping} 
Figure~\ref{fig_diff_mapping} summarizes the pictorial illustration of different mapping schemes and their characteristics. 
%\sm{Make the mapping scheme order as single group, multi group, dynamic multi group, bin, dynamic bin. That way reviewers can't complain if bin mapping is the simplest one why we tried other mapping schemes. Rather we can portray the story as we start with single group, move to multi group then find bin mapping and turns out bin mapping is better than others.}

% --> figure mapping all in one plot

\paragraph{Single Group Mapping:}
\label{sec-single-group}

In this design, we choose $\vert \vocab \vert$ functions to express an $l$-length program, $P$.
Each function represents a DSL token and takes one continuous parameter. Each function will map its parameter to at most one DSL token. Thus,
$P=\left<f_1(x^1_1), f_2(x^2_1), \dots, f_{|\vocab|}(x^{|\vocab|}_1)\right>$ where $f_i(x^i_1):\mathbb{R}\rightarrow\vocab^{0|1}$.
%variables for \acro\, where each variable represents a token (i.e., DSL function) in $\vocab$.
%First variable of \acro\ represents DSL's first function and so on. Here all the variables value are also real number. \acro\ tries to optimize the value. 
\schemegen\ samples each continuous parameter and then constructs the program by choosing the tokens corresponding to the $l$ largest sampled values.
%with the largest values.  
The token for the largest sampled value occurs first in the program.  The token corresponding to the next largest value occurs second, and so on. This is shown in Figure 2(a). %Figure~\ref{fig_single_group}.
%Top variable means they have higher numerical value than others. As we mapped $M$ variables from \acro\ to $M$ functions of our DSL, we can easily construct the program from top $N$ chosen variables. The program construction idea is to put first top function to the first place, second top function to the second place and so on. After constructing the full $N$ length program we can evaluate that and get fitness score.
A consequence of this approach is that each DSL token could occur at most once in a program. Moreover, the mapping scheme is not applicable if the length of the program, $l$ is larger than $|\vocab|$. The formal definition of $f_i$ and the corresponding selection of program token is as follows:
\begin{align*}
f_i(x^i_1)&=\begin{cases}
D_i & \mathrm{if}\, x^i_1 = Top_j\left(\{x^i_1\}^{|\vocab|}_{i=1}\right) \,\,\, \mathrm{for}\,\, 1\le j\le l \\
None & otherwise
\end{cases} \\
P_j &=D_i \quad\qquad when ~f_i(x^i_1)=D_i,
\end{align*}
%\begin{align*}
%\begin{tabular}{lll}
%\multirow{2}{*}{$f_i(x^i_1)=$}& $D_i$ & $if ~x^i_1 = Top_j(\{x^i_1\}^{|\vocab|}_{i=1})$ for $1\le j\le l$\\
%& $ None$ & $otherwise$ \\
%$P_j = $ & $D_i$ & $when ~f_i(x^i_1)=D_i$,    
%\end{tabular}
%s\end{align*}
where $Top_j(ST)$ indicates the $j$-th top element of the set, $ST$. $None$ indicates that $f_i$ does not map its parameter to any DSL token. This scheme is a very basic mapping scheme.
%among other mapping schemes. 
Here a DSL token can not be used more than once while constructing a program. 
%It is not a limitation rather an artifact of this mapping scheme.
%In other words, the range of $f_i$ becomes $\vocab^0$ or $\emptyset$.

%\taa{This probably isn't a problem for the results here because our target programs don't have duplicates either but would be a big problem in a practical system.}
%\taa{There's probably a mathematical representation of this process we should put here instead of just a description.}

% \begin{figure}[hptb]
% \begin{center}
% \includegraphics[width=0.35\columnwidth]{./FIGS/multi-group}
% %\vspace{-0.2cm}
% \caption{Representation of multi-group mapping. }
% %\jt{R and D have a different math font than the text in these figures 
% % $D^{0|1}$ has 2 options in the right ellipse. The second drawing appears like there were 3. Also, like in figure 1, shall we add labels for x1..xL and DSL tokens?}}
% \label{fig_multi_group}
% \end{center}
%  %\vspace{-0.6cm}
% \end{figure}

\paragraph{Multi-Group Mapping:}
\label{sec-multigroup}
In this mapping, \schemegen\ uses $l$ functions, each taking $|\vocab|$ continuous parameters to express the program $P$.  Each position in the program is thus represented by one function. Each parameter of that function corresponds to a DSL token.
%for each DSL function.
\schemegen\ samples the continuous parameters of the first function and selects the largest sampled parameter. The corresponding DSL token becomes the first token in $P$. Then, \schemegen\ samples the parameters of the second function and the token corresponding to the largest sampled parameter becomes the second token in $P$. This process repeats for each of the $l$ functions. This is shown in Figure 2(b). %~\ref{fig_multi_group}. 
The formal definition of $f_i$ and the corresponding selection of program token is as follows:
%\begin{table}[h]
%\centering
%\scalebox{1}{
\begin{align*}
f_i(x^i_1,\dots, x^i_{|\vocab|}) &= D_j \quad \mathrm{if} \,\, x^i_j = Top_1\left(\{x^i_j\}^{|\vocab|}_{j=1}\right) \\
P_i &= D_j.
\end{align*}

This mapping scheme is flexible in the sense it allows a DSL token to be used multiple times in a program. 
%With this mapping scheme a program can be constructed with the same DSL token multiple times in any place of the program. 
Thus, it is more robust than Single Group Mapping scheme but it requires a larger number of variables to represent a program.

%}
%\end{table}
%, we select the sample from the first group of $\vert \vocab \vert$ variables with the largest value and the corresponding token becomes the first entry in the program.  Then, the sample values of the next group of $\vert \vocab \vert$ variables are examined and the token corresponding to the largest sample value becomes the second entry.  This process repeats a total of $N$ times, once for each position in the program.
%This idea is same as single group ranking. But instead of taking all the functions from same group we construct $N$ group for $N$ length program. Each of the group has $M$ variables each were $M$ represents the function from our DSL. Now while constructing the program we look into each group and see which variable has the top value. As we already know which variable represent which functions from our DSL we can choose one function from each group and construct the program. In this way each of the gene in \scheme has $M*N$ variables.

% \begin{figure}[hptb]
% \begin{center}
% \includegraphics[width=0.35\columnwidth]{./FIGS/dynamic-multi-group}
% %\vspace{-0.2cm}
% \caption{Representation of dynamic multi-group mapping. }
% %\jt{R and D have a different math font than the text in these figures 
% % $D^{0|1}$ has 2 options in the right ellipse. The second drawing appears like there were 3. Also, like in figure 1, shall we add labels for x1..xL and DSL tokens?}}
% \label{fig_dyn_multi_group}
% \end{center}
%  %\vspace{-0.6cm}
% \end{figure}

\paragraph{Dynamic Multi-Group Mapping:}
\label{sec-dynamic-multi-group}
In multi-group mapping, we have $l$ functions, each taking $|\vocab|$ continuous parameters and choosing one DSL token for $P$, whereas in single group mapping, we have $|\vocab|$ functions, each taking one continuous parameter and choosing at most one DSL token for $P$.
%function from each of the group. 
Dynamic multi-group mapping is a hybrid scheme between single and multi-group mapping. Here, we have $k$ functions, each taking $|\vocab|$ continuous parameters and choosing $\nicefrac{l}{k}$ DSL tokens for $P$. \schemegen\ randomly chooses $k$ for each program. Each function uses single group mapping scheme to choose $\nicefrac{l}{k}$ DSL tokens. This is shown in Figure 2(c). %~\ref{fig_dyn_multi_group}. 
This mapping gives \schemegen\ the flexibility to find out the best group setup by evolving. The formal definition of $f_i(\cdot)$ and the corresponding selection of program token is as follows:
% \begin{align*}  % this equation does not fit into two column 
% %\begin{table}[h]
% %\centering
% %\scalebox{1}{
% % \begin{tabular}{lll}
% f_i(x^i_1,\dots, x^i_{|\vocab|}) &= \left\{D_j | ~x^i_j = Top_{\frac{l}{k}}\left(\{x^i_j\}^{|\vocab|}_{j=1}\right)\right\}\\
% P_{(i-1)\frac{l}{k}+t} &= D_j \quad \mathrm{if}\,\, x^i_j = Top_t\left(\{x^i_j\}^{|\vocab|}_{j=1}\right) \quad \mathrm{for}\,\, 1\le t\le \frac{l}{k}.
% % \end{tabular}
% %}
% %\end{table}
% \end{align*}
\begin{align*}  
%\begin{table}[h]
%\centering
%\scalebox{1}{
% \begin{tabular}{lll}
% use \\ then in next line & to break down equation
f_i(x^i_1,\dots, x^i_{|\vocab|}) &= \left\{D_j | ~x^i_j = Top_{\frac{l}{k}}\left(\{x^i_j\}^{|\vocab|}_{j=1}\right)\right\}\\
P_{(i-1)\frac{l}{k}+t} &= D_j \quad \mathrm{if}\,\, x^i_j = Top_t\left(\{x^i_j\}^{|\vocab|}_{j=1}\right) \quad \mathrm{for}\,\, 1\le t\le \frac{l}{k}.
% \end{tabular}
%}
%\end{table}
\end{align*}

Here, \schemegen\ can decide the number of groups to construct a program. Although \schemegen\ can decide the number of groups, it always requires same number of variables as Multi-Group Mapping in addition to the group variable. Hence, this schemes requires more variables than any other schemes.
%the most number of variables compare to any other schemes.

%Instead of taking one variable from each of the group, we introduce another variable $K$ in \acro\ that decide how many groups to use.

%Then we choose equal number of functions, $f$ from each $K$ groups that are necessary to construct the program. Here we use the single group ranking idea to choose $f$ functions from each of group. In this way we give \acro\ the flexibility to find out the best group setup by evolving.

% \begin{figure}[hptb]
% \begin{center}
% \includegraphics[width=0.35\columnwidth]{./FIGS/bin}
% %\vspace{-0.2cm}
% \caption{Representation of bin mapping. }
% %\jt{R and D have a different math font than the text in these figures 
% % $D^{0|1}$ has 2 options in the right ellipse. The second drawing appears like there were 3. Also, like in figure 1, shall we add labels for x1..xL and DSL tokens?}}
% \label{fig_bin_mapping}
% \end{center}
%  %\vspace{-0.6cm}
% \end{figure}

\paragraph{Bin Mapping:}
\label{sec-bin-map}
%\taa{I think this part is redundant.  All the variables in \acro\ are real values. Original purpose of \acro\ is to optimize those values in such a way that it can achieve maximum fitness score. We use this idea into binning mechanism to construct program for program synthesis.} 
In bin mapping, we choose $l$ functions to
%continuous random variables to 
construct an $l$-length program $P$. Each function takes one continuous parameter. Thus, 
$P=\left<f_1(x^1_1), f_2(x^2_1), \dots, f_l(x^l_1)\right>$ where $f_i(x^i_1):\mathbb{R}\rightarrow\vocab$.
%But the main challenge is to construct such a mechanism that can tell us for which value represent which function as the values are all real numbers. To solve this problem we get the idea of using binning. 
Each function $f_i$ must be able to map the continuous parameter to one of the DSL tokens.
This is shown in Figure 2(d). %~\ref{fig_bin_mapping}.
%Each such variable must be able to map onto each of tokens in $\vocab$.  
%To do this, 
We conceptually divide the range of each variable 
%\taa{we abandoned variable sized binning right?} equally into 
into $\vert \vocab \vert$ bins.
%(i.e., the number of tokens in our DSL). 
The size of each bin can be equal (except the bin at each end) or proportional to the probability of the corresponding token being present in the program $P$. 
%\jt{Bins cannot be equally sized because the domain is R and two bins (in the extremes) should be bigger. If we assume $0 \le x_i \le 1$, then we should refer to the $0\dots 1$ domain instead of R.} 
Prior work~\cite{netsyn, deepcoder} showed how to infer such probability. 
When \schemegen\ samples from the continuous variables to create a candidate program, it takes each real sampled value and determines into which of the bins the value falls.  The token used in that position in the candidate program is the one corresponding to the bin number into which the sampled value fell.  This sampling process occurs for each of the $l$ continuous parameters.
%until the $l$ length program is constructed. 
Thus, the definition of $f_i$ and the corresponding selection of program token is as follows: %\vspace{-0.3cm}
\begin{align*}
f_i(x^i_1) & =D_j \qquad \mathrm{if}~x^i_1\in Bin_{D_j} ~\mathrm{for}~1\le j\le |\vocab| \\
P_i & = D_j, 
\end{align*}
%\begin{table}[h]
%\centering
%\scalebox{1}{
% \begin{align*}
% \begin{tabular}{lll}
% $f_i(x^i_1)=$ & $D_j$ & $if~x^i_1\in Bin_{D_j} ~for~1\le j\le |\vocab|$\\
% $P_i=$ & $D_j$ &\\
% \end{tabular}
% \end{align*}
%}
%\vspace{-0.3cm}
%\end{table}
% \begin{align*}
% f_i(x^i_1) & =D_j \qquad \mathrm{if}~x^i_1\in Bin_{D_j} ~\mathrm{for}~1\le j\le |\vocab| \nonumber\\
% P_i & = D_j,
% \end{align*}
where $Bin_{D_j}$ represents the range of values corresponding to token $D_j$, and $P_i$ represents the $i$-th token of program $P$. 
This mapping scheme is more robust and flexible compared to 
the other mapping schemes. %Compare to other schemes, 
It requires the less number of variables to represent a program. This makes the optimization problem easier and helps \schemegen\ to find more programs within a fixed time budget.
%restriction compare to other schemes.

%\taa{There's probably a mathematical representation of this process we should put here instead of just a description.}
%evolve with it's own mechanism. The idea of binning comes handy when we construct the program out of the \acro\ gene. When constructing the program we see in which bin one variable's value reside into. All the bins are mapped to different functions from the DSL. Thus, we can choose function for each of the \acro\ variables and construct a $N$ length program. We can evaluate the program by our usual way. When evaluating we also get it's fitness score from the fitness function and assign to that particular gene.

% \begin{figure}[hptb]
% \begin{center}
% \includegraphics[width=0.35\columnwidth]{./FIGS/dynamic-bin}
% %\vspace{-0.2cm}
% \caption{Representation of dynamic bin mapping. }
% %\jt{R and D have a different math font than the text in these figures 
% % $D^{0|1}$ has 2 options in the right ellipse. The second drawing appears like there were 3. Also, like in figure 1, shall we add labels for x1..xL and DSL tokens?}}
% \label{fig_dyn_bin_mapping}
% \end{center}
%  %\vspace{-0.6cm}
% \end{figure}

\paragraph{Dynamic Bin Mapping:}
\label{sec-dynamic-bin}
With prior mapping schemes, \schemegen\ can synthesize programs of a particular length. However, it is possible to find a program of a lower length (than the length of the target program) that also satisfies the given specification. To accommodate such programs, we propose dynamic bin mapping scheme which uses one function to choose the length and other functions to choose the DSL tokens. Thus, we have $l+1$ functions each taking one continuous parameter. Function $f_{l+1}$ maps its parameter to a program length at which \schemegen\ will synthesize a program. To do this, we conceptually divide the range of $x_{l+1}^0$ into $l$ equal sized bins. When \schemegen\ samples this parameter, it determines which bin the parameter's value fall into. The length corresponding to that bin is chosen as the program length. Earlier functions (up to that length) map their continuous parameters into one DSL token each based on the bin mapping idea. This is shown in Figure 2(e). %~\ref{fig_dyn_bin_mapping}. 
The formal definition is as follows:  
\begin{align*}
f_i(x^i_1) & =D_j \qquad \mathrm{if}~x^i_1\in Bin_{D_j} ~\mathrm{and}~x^{l+1}_1\in Bin_k\\ 
& \qquad\qquad\mathrm{for}~1\le j\le |\vocab| ~\mathrm{and}~1\le k\le l ~\mathrm{and}~i\le k\\
P_i & = D_j, 
\end{align*}

%Pros of this mapping scheme is, i
This mapping scheme can represent variable length program instead of always constructing the highest length program. However, 
%due to the error in selecting the target program length, 
synthesis rate can be lower if length is not accurate.

\subsubsection{Restart Policy}
\label{sec-restart-policy}

\schemegen\ can stop its evolutionary process when
%for multiple reasons.  First, it may stop when it reaches a maximum user-defined threshold for the number of %generations. Second, it may also stop when it detects that 
the potential solutions are converging due to some local minima.
%\taa{need the technical description of this detection here}. 
This is checked by determining if a change in one or multiple axes does not affect the distribution mean, or if the condition number of the covariance matrix is too high (i.e., ill-conditioned), or the evolutionary path's step size is too small to reach the solution~\cite{cmaes.hansen}. 
%\jt{I don't understand what you mean with ``condition of the matrix''. How do we measure it? Are we referring to the condition number of a matrix? If not, can we use math to define it?}
%When \scheme\ converges without finding an equivalent program, it is because it has converged to a local minima. % 
To escape from such situation, we investigate several restart policies where \schemegen\ restarts with fresh initial parameters.
%in those cases where \acro\ stops before reaching the time threshold.  
We explore all policies resulting from the combinations of 3 core restart policies: population-based (PB), mean-based (MB), and covariance matrix-based (CB).
For the PB restart, \schemegen\ doubles the size of the population from its previous size. 
%For PB restart, \scheme\ re-initializes every parameter of CMA-ES except that the population size is doubled from the previous value. 
%not resetting the \acro\ mean vector, and not resetting the \acro\ covariance matrix.  For example, at each restart, one could double the population size.  
%Initially, \scheme\ starts with a low population size because the run time depends on the population size. Namely, a smaller population size allows \scheme\ to find programs quickly. 
%\jt{Are we expecting our method to be way slower? How much? Are we setting a limit to PB (for example, max is 8 times the initial size to make sure it is not too slow after some time)?}
%\scheme\ doubles the population size at every restart under the PB restart policy. \jt{This repeats the previous sentences}
For the MB restart policy, \schemegen\ resets the mean vector to randomly initialized values
%its default value 
(using the uniform random distribution) after a restart instead of keeping its current value. % from the previous iteration. %\jt{What is the ``default value''? Do we draw a new mean vector or do we store the initial one and reuse it? I'm not sure that I understood what we do as the next sentence is also confusing (we keep the vector?).}
%For MB restart policy, \scheme\ re-initializes every parameter of CMA-ES except the mean vector, which maintains its latest content. %The intuition for keeping the mean vector as opposed to re-initializing it is that \scheme\ retains some knowledge about different parameters' individual values across restarts and may therefore converge faster after a restart. 
For the CB restart policy, \schemegen\ re-initializes the covariance matrix to the identity matrix after a restart instead of keeping its current values. 
%For the CB restart policy, \scheme\ re-initializes every parameter except the covariance matrix that remains unchanged. 
%Here, the intuition for keeping the covariance matrix as opposed to re-initializing it is that \scheme\ retains its knowledge about the dependencies among the parameters, and may able to converge faster after a restart. %\jt{Same as MB, are we resetting it or keeping it after a restart? Here, writing is confusing.}
%\jt{For MB and CB we don't say to what exactly the mean and the covariance are initialized/restarted to. If this is given in Section 2, then we should mention it.}
Additional restart policies can be constructed by any combination of these core restart policies. Some of these combinations have been proposed in earlier work such as IPOP~\cite{ipop} and BIPOP~\cite{bipop} where they re-initialize mean vector, covariance matrix, and increase population size simultaneously.
%\jt{MISSING: we should refer to previous work and give references to existing restart policies.}

%In case of PB restart,  population size restart technique, we double the previous population size each time we restart. The second restart technique is to not reset the mean vector of \acro\ variables which otherwise would be initialized randomly.  The third technique is to not reset the covariance matrix, which would otherwise be set to the identity matrix.  The intuition is that the restarted \acro\ may converge faster by retaining some information learned during a previous attempt but these could also immediately make it more like \acro\ gets stuck in the same local maxima.

%\taa{This goes in results section} We try these three restart policy and empirically shows that combining these restart policy together increase program synthesis rate.

\begin{figure*}%[!htbp]
 %   \vskip 0.2in
    \begin{center}
        %\centerline{\includegraphics[trim=0 10 0 230, clip, width=0.8\textwidth]{./FIGS/Overview_new}}
        % \includegraphics[width=\columnwidth]{./FIGS/overview.pdf}
        % \includegraphics[width=1.4\columnwidth]{./FIGS/overview.pdf}
        \includegraphics[width=1\textwidth]{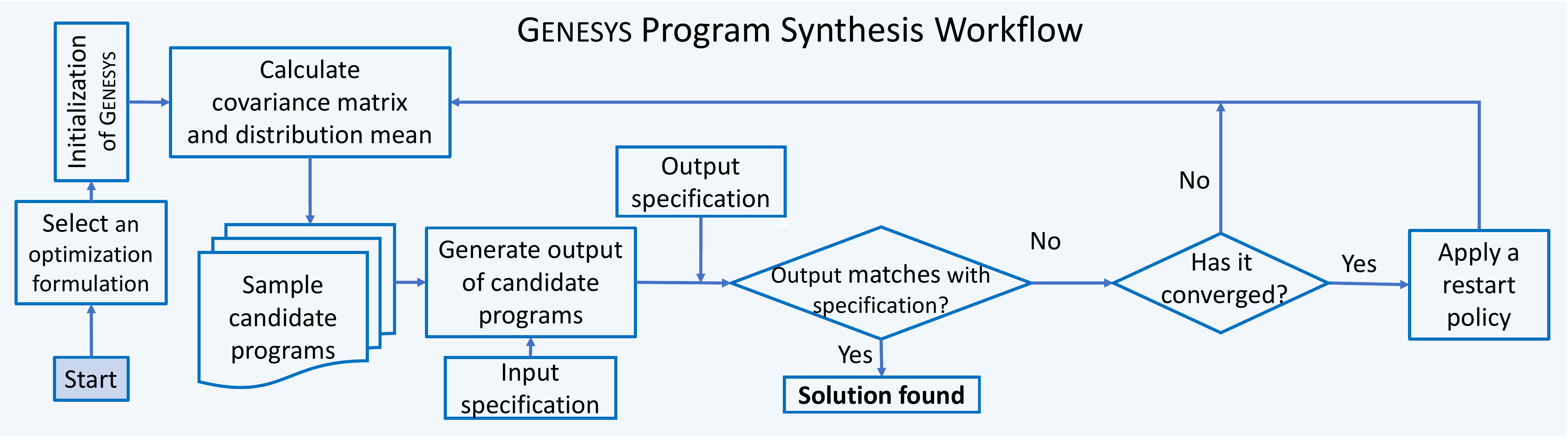}
        \caption{Overview of Genesys.}
        \label{fig-overview-genesys}
    \end{center}
    \vspace{-1.0cm}
\end{figure*}

\subsubsection{Putting It All Together}
%Description of the cmaes search process. Description of different variable mean, covariance matrix, population size.
\schemegen\ follows the high level workflow shown in Figure~\ref{fig-overview-genesys}. \schemegen\ starts by initializing a multivariate normal distribution and the population size. The continuous variables of the distribution are determined based on the choice of a mapping scheme. Any of the mapping schemes in Section~\ref{sec-sub-setup} can be used. In each generation, \schemegen\ samples $\lambda$ points from the multivariate distribution and maps each sample to a program. Each of these candidate programs could potentially be the solution. Therefore, \schemegen\ applies the error function $\mathbb{E}$. The error function essentially applies inputs from the given specification to a candidate program and checks if the produced output matches with the given output specification. If the outputs match, the candidate program is returned as the solution. Otherwise, \schemegen\ checks if convergence has reached. If not, \schemegen\ updates the mean, covariance matrix and distribution path based on the error function and starts a new generation. If, on the other hand, convergence is reached, \schemegen\ applies a restart policy and repeats the whole process with new parameters. The synthesis process continues until a maximum time limit has passed or a solution is found.  
% ----------------------- Design End ------------------

\section{Task 2: Automatic Configuration Synthesis}

In Task 2, we propose to extract configuration constraints for software automatically from textual flies such as manuals, readme files, online blogs, etc.

\paragraph{Background and motivation:} Large scale software synthesis is very challenging. One needs to tackle many small problems to synthesis a large scale software. Configuration synthesis is a big part of it. Configuration, also known as specification, defines how a software will be set up and run in the system. Mis-configuration of a software can lead to failures during runtime. It makes a system unreliable and often, difficult to maintain. Thus configuring the software with correct parameters is very important. 

Now a days, almost all large scale software comes with many configuration parameters. As a result, manually setting up those configurations can be error-prone. An automatic system needs to be placed that prevents mis-configuration. Usually these configuration constraints are written in large scale software manuals. Sometime there can also be found in input files, special configuration file, system environment file, online blogs, etc. Going through all these files automatically and extracting specifications is very challenging. In this work, we explore automatic configuration extraction from different unstructured software manuals. We refer to this work as {\it ConfigSyn}.

\begin{table}[hbtp]
\centering
\scalebox{1}{
\begin{tabular}{lll}
\toprule
Category        & Specification & Example\\
\midrule
Value           & $p==v$              & Set this value greater than 2000\\
                & $p < v  |  p > v$   &  \\
                & $p\in \{v, v'\}$    &  \\
\midrule
Correlation     & $with(p, p')$       &  $p$ lower than $p'$ not ideal \\
\midrule 
Usage           & $use(p)$            & this option is useful for diagnostic purpose \\
\midrule 
Property        & $format(p, f)$      & best to specify $p$ as an absolute path\\
\bottomrule
\end{tabular}}
\caption{Category of specifications and their examples.}
  \label{table:spec_category}
%\vspace{-0.6cm}
\end{table}

\paragraph{Example of specifications:}
We categorize software configuration specification into 4 categories. Table \ref{table:spec_category} shows the category with specification and their examples. This specifications can be written in software manuals or any online group. For example, let us assume that specification $p$ needs to be within two value $on$ or $off$. If someone configures $p$ with a value different from $on, off$, the software will be mis-configured that can lead to a crash. The goal of our proposed {\it ConfigSyn} is to collect the specification of $p$ by analyzing the software manuals or other input sources. And this specification can be used as a rule checker during the configuration of the software. 

\begin{table}[hbtp]
\centering
\scalebox{1}{
\begin{tabular}{lc}
\toprule
Software        & Number of Configuration Parameter\\
\midrule
Mysql           & 1453  \\
\midrule
Httpd     & 372  \\
\midrule 
Hadoop           & 132             \\
\midrule 
CentOs        & 407      \\
\bottomrule
\end{tabular}}
\caption{Number of configuration parameters of different software.}
  \label{table:no_config_param}
% \vspace{-0.6cm}
\end{table}

\paragraph{Data collection:}
We choose to synthesize configurations for Mysql, Httpd, Hadoop, and CentOS. To do this, we first collect the software manual for these four software. We go through the manual and collect all the configuration parameters. Table \ref{table:no_config_param} shows the number of configuration parameters for each of the software. Then, we collect configuration-related raw data set mainly from two sources. First, we go through each of the sentences of the manual and if any of the configuration keyword has been found we collect that sentence assuming this sentence may contain specifications related to this parameter. Then, we build a crawler to go through different Stackoverflow\cite{stackoverflow} posts. Stackoverflow is a developer website where people submit different questions related to software and other people answer those questions. It has a lot of posts related to configuration bugs and their solutions. The crawler collects those posts and makes a database. This database contains raw texts that hold different specifications of the configurations. As these texts are all written in a natural language (e.g., English), we use a deep learning model to extract the specifications automatically.

\paragraph{Model:}

%We formulate our problem into a named-entity-recognition (NER) \cite{NER} problem. NER is a well studied problem in natural language processing. In NER method, model detects different types of named entity in natural text. 

We formulate this problem as a sequence-to-sequence translation problem. For specification modeling, we use a transformer-based~\cite{transformer} end-to-end encoder-decoder model to synthesize configuration specifications. As the configuration is described in a natural language, we first encode the natural texts with an encoder into a latent representation. Then, on the decoder side if any configuration is present in the texts, that is decoded automatically. For example, sentence {\it set the \detokenize{port_num} to any value greater than 2000} has configuration specification as {\it \detokenize{port_num} > 2000}. Now, we can train a model which encodes the sentence {\it set the \detokenize{port_num} to any value greater than 2000} into some latent representation and the decoder will take it and decode to {\it \detokenize{port_num} > 2000}. 

\paragraph{Transformer model:} For the core modeling of our dataset, we use the Transformer-based~\cite{transformer} sequence-to-sequence model. This model achieves the state-of-the-art performance on various machine translation-based work. It significantly outperforms other models (such as RNN, LSTM, etc.) that rely on recurrence-based calculations. The transformer based model is an encoder-decoder based model that harnesses the power of attention mechanism. There are two parts in transformer-based model. One is encoder and another is decoder. First the input goes through the embedding layer to make token embedding for each of the input tokens. Positions of each token are represented by positional embedding. Token embedding and positional embedding are merged together and go through multi-head attention mechanism. Attention mechanism finds correlation between each of the tokens that are passed to the encoder. The purpose of the encoder is to convert the inputs into some latent representation. This latent representation is passed to the decoder part of the transformer model. Decoder takes the latent representation from the encoder and a start symbol. It continuously pops out an output token until it sees the end token or maximum length is reached. Basically, the decoder gives all tokens' probabilities through the softmax layer. The token with the highest probability is taken as the output token.  

% \paragraph{Evaluation of extracted configuration:}

%\paragraph{Research Plan:}

% We propose to model configuration synthesis as an information retrieval problem from natural language. Extracting information such as named entity (NER)\cite{??} from natural language is a well studied problem. Scientific community are using state-of-the-art pre-trained natural language embedded model to extract various named entity from textual data. We model different types of software configuration as named entity and extract those with a pre-trained language model. The main challenge of {\it ConfigSyn} is to create a model that is unbiased and generalized. To achieve this we need uniformly distributed training data. Also, training any machine learning model needs large amount of data. There is no public dataset readily available for configuration synthesis. Also, the process can not be automated. Thus, we need to create the dataset manually by tagging and processing by human. After collecting the data we can train model and design the framework for configuration synthesis. Thus, the plan to achieve this goal is as follows:

\section{Related work}
% start of related work grabbed from NetSyn paper
Program synthesis has been widely studied with various applications. \emph{Formal program synthesis} uses formal and rule based methods to generate programs\cite{mana75, wang2018relational}. Formal program synthesis usually guarantees some program properties by evaluating a generated program's semantics against a corresponding specification~\cite{gulwani12, alur15}. In ~\cite{feng18}, authors use a SMT based solver to find different constraints in the program and learn useful lemma that helps to prune away large parts of the search space to synthesize programs faster. However, these techniques can often be limited by exponentially increasing computational overhead that grows with the program's instruction size~\cite{Heule:2016:pldi, Bodik:2013:sttt, Solar-Lezama:2006:asplos, Loncaric:2018:icse, Cheung:2012:cikm}.

% MP with ML - probabilistic correctness
Another way to formal methods for program synthesis is to use machine learning (ML). Machine learning differs from traditional formal program synthesis in that it generally does not provide correctness guarantees. Instead, ML-driven program synthesis approaches are usually only \emph{probabilistically} correct, i.e., their results are derived from sample data relying on statistical significance~\cite{Murphy:2012:mitpress}. Such ML approaches tend to explore software program generation using an objective function. Objective functions are used to guide an ML system's exploration of a problem space to find a solution. Other deep learning based program synthesizer ~\cite{bunel2018leveraging, parisotto2016neuro, chen2020program} also tried different approaches such as reinforcement learning to solve the problem. Most of these works focus on synthesizing programs in domain specific languages, such as FlashFill~\cite{robustfill, kalyan2018neural} for string transformation problem, simulated robot navigation, such as Karel~\cite{shin2018improving, chen2018execution} or LIST manipulation~\cite{netsyn, deepcoder, polosukhin2018neural, nye2019learning, Zohar:2018:nips, feng18} work.

%An alternative to formal methods for MP is to use machine learning (ML)~\cite{deepcoder, raychev14, bunel18, npi, cai17, aip, real17, real18}. Machine learning differs from traditional formal program synthesis in that it generally does not provide correctness guarantees. Instead, ML-driven MP approaches are usually only \emph{probabilistically} correct, i.e., their results are derived from sample data relying on statistical significance~\cite{Murphy:2012:mitpress}. Such ML approaches tend to explore software program generation using an objective function. Objective functions are used to guide an ML system's exploration of a problem space to find a solution. %For example, \emph{backpropagation}, the objective function used in deep neural networks (DNNs), is a gradient-based optimization algorithm that aims to iteratively reduce the error of the DNN using mathematical differentiation~\cite{Rumelhart:1986:nature}. 

% Prior work on ML for MP
%\jt{Please can someone confirm that the citations in this paragraph are NN-based? A few of these, like Real et al. are cited as GA methods as well.}
Among the ML-based program synthesis, in Deepcoder\cite{deepcoder}, the authors train a neural network with input-output examples to predict the probabilities of the functions that are most likely to be used in a program. Raychev et al.~\cite{raychev14} take a different approach and use an n-gram model %(a technique that analyzes a contiguous sequence of elements) 
to predict the functions that are most likely to complete a partially constructed program.
Robustfill~\cite{robustfill} encodes input-output examples using a series of 
recurrent neural networks (RNN), and generates the the program using another RNN one token at a time.
Bunel et al.~\cite{bunel18} explore a unique approach that combines reinforcement learning (RL) with a supervised model to find semantically correct programs. These are only a few of the works in the program synthesis space using neural networks~\cite{netsyn, dilling-multi_spec, npi, cai17, Chen18}.

%As demonstrated by Becker and Gottschlich~\cite{aip}, genetic algorithms also show promise for MP. 
Significant research has been done in the field of genetic programming~\cite{stackgp, lgp, allgp} too, where the goal is to find a solution in the form of a complete or partial program for a given specification. Prior work in this field has tended to focus on either the representation of programs or operators during the evolution process. 
Real et al.~\cite{real18} recently demonstrated that genetic algorithms can generate accurate image classifiers. Their approach produced a state-of-the-art classifier for CIFAR-10~\cite{cifar10} and ImageNet~\cite{imagenet} datasets. Moreover, genetic algorithms have been exploited to successfully automate the neural architecture optimization process~\cite{salimans17, such17, liu17, santientlabs, real2020automlzero}.

However, all these works formulate program synthesis as a search problem of discrete parameters. On the other hand, \scheme\ tries to formulate program synthesis as a continuous optimization problem and uses a well established derivative free method, CMA-ES, to solve it. 
%To our knowledge, there is no known work with \acro\ in the field of machine programming. However, 
Previously, CMA-ES was used in machine learning~\cite{mueller1999a}, 
aerospace engineering~\cite{Lutz1997}, mechanical engineering~\cite{Sonoda2004}, health~\cite{Winter2002} and various other engineering fields~\cite{cma-es-app} but not in the field of synthesizing complex programs.
%\jt{Cite NPO and don't claim to be first in the continuous domain - per our last meeting discussion.}

In addition, researchers engage in extensive analysis of configuration files in order to mitigate the occurrence of software misconfigurations. The prevention of such misconfigurations holds various significances for ensuring software reliability. Therefore, researchers dedicate their efforts towards identifying configurations valid rules and bounds (e.g., specifications), with the aim of preventing software misconfigurations~\cite{specsyn, configV, nadi.icse.2014}.

\section{Conclusion}
We propose automatic synthesis of software code and configuration constraints as two major tasks for the automation of software generation process. We sub divide the software synthesis task into two major sub-tasks that explore the synthesis with discrete and continuous search techniques. On the other hand, we propose to use a state-of-the-art natural language model for configuration specification extraction. We believe our approaches in each of the tasks open up potential other research directions to explore.

\bibliographystyle{unsrt}
\bibliography{main}

\end{document}